\documentclass[12pt]{iopart}

\usepackage{graphicx}

\begin{document}

\title[Simulations of ion beam instabilities]{Two-dimensional PIC 
simulations of ion-beam instabilities in Supernova-driven plasma flows}

\author{M E Dieckmann\textsuperscript{1,2}, A Meli\textsuperscript{3}, P K Shukla\textsuperscript{1}, L O C Drury\textsuperscript{2} and A Mastichiadis\textsuperscript{3}}

\address{1 Institut f\"ur Theoretische Physik IV, Ruhr-Universit\"at Bochum, D-44780 Bochum, Germany}
\address{2 Dublin Institute for Advanced Studies, Dublin 2, Ireland}
\address{3 Department of Physics, National University of Athens, Panepistimiopolis, Zografos 15783, Greece}
\ead{markd@tp4.rub.de}

\begin{abstract}
Supernova remnant (SNR) blast shells can reach the flow speed $v_s = 0.1 \,c$ and shocks form at its front. Instabilities driven by shock-reflected ion beams heat the plasma in the foreshock, which may inject particles into diffusive acceleration. The ion beams can have the speed $v_b \approx v_s$. For $v_b \ll v_s$ the Buneman or upper-hybrid instabilities dominate, while for $v_b \gg v_s$ the filamentation and mixed modes grow faster. Here the relevant waves for $v_b \approx v_s$ are examined and how they interact nonlinearly with the particles. The collision of two plasma clouds at the speed $v_s$ is modelled with particle-in-cell (PIC) simulations, which convect with them magnetic fields oriented perpendicular to their flow velocity vector. One simulation models equally dense clouds and the other one uses a density ratio of 2. Both simulations show upper-hybrid waves that are planar over large spatial intervals and that accelerate electrons to $\sim$ 10 keV. The symmetric collision yields only short oscillatory wave pulses, while the asymmetric collision also produces large-scale electric fields, probably through a magnetic pressure gradient. The large-scale fields destroy the electron phase space holes and they accelerate the ions, which facilitates the formation of a precursor shock.\end{abstract}

\pacs{98.38.Mz, 52.35.Qz, 52.65.Rr}
\maketitle

\section{Introduction}

It is well-established that a supernova remnant (SNR) can accelerate particles to energies $\sim 10^{15}$ eV, probably through diffusive acceleration [1-3] by shocks between the SNR blast shell and the interstellar medium. In particular perpendicular shocks, for which the ambient magnetic field vector is orthogonal to the shock normal, are likely to be efficient particle (i.e. electron or proton) Fermi accelerators \cite{MeliBierm06}. Diffusive acceleration across perpendicular shocks requires that the particle gyroradius exceeds the shock thickness. In the case of the electrons, suprathermal populations with energies in excess of 100 keV \cite{Paper4} must be present to start with the acceleration. This is the so-called 'injection problem'. The injection of electrons is probably achieved by waves in the foreshock and by the shock precursors [6-13]. 

Shocks are sources of ion beams, which originate from upstream ions that have been reflected by the shock or from downstream ions that leaked through the shock \cite{Paper4,Foreshock,Malkov}. The spectrum of beam-driven waves depends on the beam speed $v_b = |\mathbf{v}_b|$ in the upstream plasma. The typical values of $v_b$ are comparable to the shock speed $v_c = |\mathbf{v}_c|$, which is $v_c \le 10^{-2} \, c$ for most solar system plasmas. The corresponding $v_b$ imply the growth of quasi-planar waves by Buneman instabilities, either if no magnetic field $\mathbf{B}_0$ is present or if $\mathbf{v}_b \parallel \mathbf{B}_0$. The Buneman instability is modified if $\mathbf{v}_b \cdot \mathbf{B}_0 \neq 0$ and upper-hybrid waves grow if $\mathbf{v}_b \perp \mathbf{B}_0$. Both waves saturate by forming electron phase space holes [16-19]. Upper-hybrid waves accelerate further the trapped electrons by their transport across the magnetic field, by which they form high-dimensional phase space structures \cite{Paper15}. Higher shock speeds of $|\mathbf{v}_c| > 0.5 \, c$ exist at some SNRs and at gamma ray bursts \cite{Kulkarni,Waxman} and ion beams moving at such speeds give rise to filamentation- and mixed mode instabilities [23-27]. Both modes saturate by forming current channels, which thermalize the plasma by their mergers. The non-linear evolution of the instability and the interaction of the waves with the particles depends on whether it is the Buneman instability, the quasi-electrostatic upper-hybrid instability or the mixed mode and filamentation instabilities with their electromagnetic component. The type of the instability thus determines the electron acceleration and injection efficiency.

Shock speeds of $v_c \approx 0.1 \, c$, which are associated with the fastest supernova-driven bulk flows \cite{Kulkarni,FryerNew}, are between both extremes and it is not a priory known which instabilities are driven by ion beams with $v_b \approx 0.1 \, c$. Wave instabilities driven by such beams have been examined with one-dimensional PIC simulations \cite{PaperNew2} and two-dimensional ones \cite{Paper17} with periodic boundary conditions, which limits the resolved wave spectrum and neglects the effects introduced by the beam front. The higher-dimensional wave spectrum driven by plasma beams with $v_b \approx 0.1c$, with a finite beam length and with a perpendicular magnetic field has remained practically unexplored, in spite of the importance of such a system for the electron injection at SNR shocks. Here we perform two simulation case studies that address such a system and we compare their results.

We perform 2(1/2)D PIC simulations of colliding plasma clouds in the $x-y$ plane, which evolve in time the three-dimensional particle momenta and electromagnetic field vectors. SNR shocks are typically magnetized and the ratio between the electron plasma frequency and the electron cyclotron frequency may be a few ten to hundred. Plasma processes close to the front of SNR blast shells can amplify the magnetic field in the upstream to give a lower ratio \cite{VolkM,BellM}. An uniform magnetic field along the z-direction is introduced as an initial condition. Its strength gives a frequency ratio of 5. This low value allows us to examine with PIC simulations the magnetic field effects \cite{Paper15}. Both colliding plasma clouds have the same density in one simulation, which gives a symmetric collision. A density ratio of 2 is employed in the second simulation. The spectrum of unstable waves is a function of the relative beam density and we can examine how robust the wave spectrum is against changes in it. The speed, with which both clouds collide along the $x$-direction, is set to $|\mathbf{v}_c| = 0.1 \, c$ and the ion beam instabilities developing out of the collision have a $v_b = v_c$. The electrons are heated and reach peak values of 20 keV in the cloud overlap region, which expands in time due to the ion motion. This value is in line with the results of previous PIC simulations of spatially homogeneous beam instabilities in the presence of a perpendicular magnetic field \cite{PaperNew2,Paper17}. 
 
The simulations show that the dominant wave structures are quasi-planar. Two-dimensional electromagnetic field structures develop well behind the ion beam front. They fluctuate and they have a lower amplitude than the planar electrostatic fields. Some structuring orthogonal to the flow velocity vector is observed in an interval close to the ion beam front, but the electric field amplitudes remain below those of the planar electrostatic fields. Non-planar structures may, however, form on larger scales \cite{Paper7}. This planarity of the wave structures demonstrates that previous 1D PIC simulations \cite{Paper24,Paper25,Paper26} should be adequate for the modelling of SNR shocks with $v_c < 0.1c$. 

The magnetic field is compressed in the overlap region of both plasma clouds. The magnetic field amplitude doubles and it is constant in the overlap region, if both clouds are equally dense. The collision of the asymmetric clouds amplifies the magnetic field close to the front of the dense ion beam to over three times the initial strength and a large-scale electrostatic field along $\mathbf{v}_c$ develops, probably through the magnetic pressure gradient. It accelerates ions in the flow direction and increases their speed by 20\%. 
 
The structure of this paper is as follows. The section 2 introduces the PIC method and the initial conditions. The section 3 presents the results of both simulations. These results are discussed in the context of supernova remnant shocks in section 4.

\section{The particle-in-cell simulation and the initial conditions}

\subsection{Initial plasma parameters}

We consider a system with the size $L_x \times L_y$ in the $x,y$-plane. Two plasma clouds collide along the $x$-direction, each consisting of the electrons and ions. Each cloud is spatially homogeneous within the interval it occupies. The plasma cloud 1 is located in the interval $-L_x / 2 < x < 0$ and it moves at the speed $v_b = 0.05c$ to increasing values of $x$. The plasma cloud 2 in the interval $0 < x < L_x / 2$ moves at the speed $-v_b$. The relative flow velocity vector is $\mathbf{v}_c = (2 v_b, 0,0)$. The boundary conditions are periodic in all directions, i.e. no further plasma enters through the boundaries.

The plasma frequency of the electrons of cloud 1 is $\omega_{e,1} = {(e^2 n_e/ \epsilon_0 m_e)}^{1/2}$, where $e$, $m_e$ and $n_e$ are the elementary charge, the electron mass and the electron number density, respectively. The ion mass $m_i = R \, m_e$ with $R = 10^3$. The ion plasma frequency of cloud 1 is $\omega_{i,1} = \omega_{e,1} R^{-1/2}$. The density of the plasma cloud 2 equals that of cloud 1 in the case study 1 (run 1) and the electron and ion plasma frequencies of cloud 2 are $\omega_{e,2} = \omega_{e,1}$ and $\omega_{i,2}=\omega_{i,1}$, respectively. In the case study 2 (run 2) we set $\omega_{e,2}=\omega_{e,1}/\sqrt{2}$ and $\omega_{i,2} = \omega_{i,1} / \sqrt{2}$. The initial temperature of all plasma species in both simulations is $T = 1.16 \times 10^5$ K (10 eV). We use the electron skin depth $\lambda_s =  c / \omega_{e,1}$ of the cloud 1 as the length scale. A spatially uniform initial magnetic field $\mathbf{B}_0 = (0, \, 0, \, B_{z,0})$ is applied in both simulations with $\omega_{ce}/\omega_{e,1} = 0.2$, where the electron gyrofrequency $\omega_{ce} = e B_{z,0}/m_e$. The cloud motion yields a convection electric field $E_c = |\mathbf{v}_b \times \mathbf{B}_0|$. The initial electric field is $\mathbf{E}_0 = (0, \, E_c, \, 0)$ for $x < 0$ and $\mathbf{E}_0 = (0, \, -E_c, \, 0)$ for $x > 0$.  The electron Debye length of the dense cloud is $\lambda_{D,1} = \lambda_s / 226$. The electron thermal gyroradius is $r_{th,e} \approx \lambda_s / 32$. The gyroradius of the ion beams is $r_{b,i}\approx 125 \lambda_s$ for a mean speed $v_b$. 

The plasma clouds interpenetrate and instabilities thermalize the electrons. After some time, the system consists of an ion beam that interacts with the beam of incoming electrons. This would be a Buneman instability in the absence of a magnetic field and an upper-hybrid instability for the $\mathbf{B}_0$ we use \cite{PaperNew2}, provided that the mixed modes and the filamentation modes do not outgrow these electrostatic waves \cite{NewJPhys}. The fastest-growing electrostatic waves have a real value of the frequency in the rest frame of the electrons that equals the electron plasma frequency (See \cite{PaperNew2} and the cited references). This would be $\omega_u = \omega_{e,1}$ in the reference frame of cloud 1 or $\omega_u = \omega_{e,2}$ in the reference frame of cloud 2. The wavenumber of the most unstable wave is $k_u = \omega_u / 2 v_b$ since the speed difference between the ion beam and the incoming electrons is $2v_b$. Its wavelength $\lambda_u = 2\pi / k_u$. Note that the initial conditions, in particular the spatially uniform distributions, are idealized. A more realistic system is discussed by Ref. \cite{Schamel2}. However, the uniform beam velocities allow for a direct comparison with previous PIC simulations of spatially unbounded beams \cite{PaperNew2}, by which similarities and differences are easily detected.

\subsection{The simulation code and resolution}

A particle-in-cell code approximates a collision-less plasma by an incompressible phase space fluid. The phase space fluid is represented by a large number of volume elements, which we call computational particles (CPs). Each CP denoted by the index $i$ of species $j$ has a charge $q_j$ and mass $m_j$. It has the same charge-to-mass ratio as the particle species it represents, e.g. $q_j/m_j = -e/m_e$ for the electrons. The absolute values of the charge and mass can be much larger. Each CP evolves in time under the influence of the collective electromagnetic fields that have been computed from all CPs. The plasma is updated by the code with the help of the Maxwell equations and the relativistic Lorentz equation of motion for each CP. The equations can be normalized (subscript $N$) by the substitutions $\mathbf{E} = c m_e \omega_{e,1} \mathbf{E}_N / e$ and $\mathbf{B} = m_e \omega_{e,1} \mathbf{B}_N / e$ for the fields. The positions and times are $\mathbf{x} = \lambda_s \mathbf{x}_N$ and $t = t_N / \omega_{e,1}$ and the differential operators $\nabla = \lambda_s^{-1} \nabla_N$ and $dt= dt_N / \omega_{e,1}$. The velocity and momentum transformations are $\mathbf{v}=c\mathbf{v}_N$ and $\mathbf{p}= c m_e \mathbf{p}_N$. The masses and charges of the individual plasma species are $m_{j} = m_e m_{N,j}$ and $q_j = e q_{N,j}$ and $\rho = n_e e \rho_N$. The normalized equations are then 

\begin{eqnarray}
\nabla_N \times \mathbf{E}_N = - \frac{\partial \mathbf{B}_N}{\partial t_N}, \,
\nabla_N \times \mathbf{B}_N = \mathbf{J}_N + \frac{\partial \mathbf{E}_N}{\partial t_N}, \\ \frac{d\mathbf{p}_{N,i}}{dt_N} = \frac{q_{N,j}}{m_{N,j}} (\mathbf{E}_N+\mathbf{v}_{N,i} \times \mathbf{B}_N), \\ \nabla_N \cdot \mathbf{E}_N = \rho_N, \, \nabla_N \cdot \mathbf{B}_N = 0.
\end{eqnarray}

The normalized fields $\mathbf{E}_N$ and $\mathbf{B}_N$ can be converted into physical units, once $\omega_{e,1}$ is given. The electric and magnetic fields are defined on a spatial grid, while the CPs follow continuous trajectories. The position of the nodes, at which the fields are defined, are placed in our code at equal distances. The considered system size is $L_x = 80 \, \lambda_s$ and $L_y = 3 \, \lambda_s$ and it is represented by $9000 \times 340$ simulation cells. Each of the four plasma species is represented by 64 particles per cell (PPC). We use normalized units and drop the subscript $N$. The time step $\Delta_t \approx 4.33 \times 10^{-3}$. The simulation time $T_t = 225$ during which each cloud propagates the distance $1270 \Delta_x$ along the x-direction. 

\section{Simulation results}

The relevant simulated field- and particle phase space distributions are examined at selected times. Changes in the $E_z$-field and in the $p_z$-components are small due to the chosen system geometry and these components are not analysed. The considered electric $E_x$ and $E_y$ components are then compared to their time-animations during $11.25<t<225$ and in the interval $y < L_y / 2$ . The movies 1 and 2 show $E_x/E_c$ and $E_y/E_c$ of the run 1 and the movies 3 and 4 show $E_x/E_c$ and $E_y/E_c$ of the run 2.  

\subsection{Collision of equally dense clouds: run 1}

\begin{figure}[t]
\begin{center}
\includegraphics[width=0.5\textwidth]{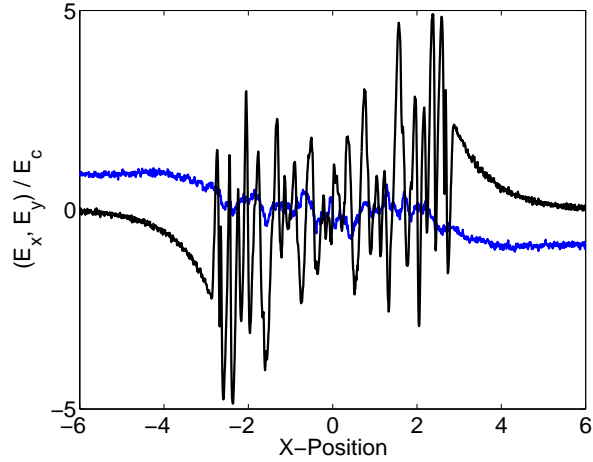}
\end{center}
\caption{The electric fields $E_x$ (black) and $E_y$ (blue) at the position $y=0$ and time $t=56$: The electric fields are practically anti-symmetric to $x=0$. The $E_x$ field is oscillatory in the interval $-3 < x < 3$, in which the clouds overlap. The $E_y$ in this interval oscillates with a low amplitude around $E_y=0$. Wave forerunners with an exponentially decreasing amplitude modulus $|E_x|$ are found for $|x|>3$.}\label{fig1}
\end{figure}

Figure \ref{fig1} displays the electric $E_x$ and $E_y$ components at $t=56$.
Both plasma clouds overlap in the interval $-2.8 < x < 2.8$, in which the $E_x$ rapidly oscillates. The oscillations of $E_y$ around $E_y = 0$ have a much lower amplitude. This is in line with what we expect from the quasi-electrostatic upper-hybrid waves, because the electrostatic wave component points along the beam velocity vector. This observation is also in agreement with the movies 1 and 2 that initially show planar wave structures, which are polarized in the x-direction. However, the observed wavelength $\lambda$ of the oscillations close to the ion beam fronts at $|x| \approx 2.8$ is less than the expected $\lambda_u \approx 0.6 \lambda_s$. 

The movie 2 reveals planar wave structures in the $E_y$-component. These waves are a consequence of the initial conditions. The abrupt change of the convective electric field at $x=0$, which is polarized along the y-direction, can couple to a wide k-spectrum of waves with $\mathbf{k}$ along the x-direction. These perturbations propagate in the fast extraordinary mode. The mean value $<E_y>=0$ in the interval $|x|< 2.8$, which is the convective electric field, evidences that there is no net flow of plasma across $\mathbf{B}_0$ in the region in which the plasma clouds overlap. This had to be expected due to the equal cloud densities. Outside $|x| < 3$ the modulus of $E_x$ decreases exponentially and the $E_y$ component converges to the value of the convective electric field.   

\begin{figure}[t]
\begin{center}
\includegraphics[width=0.5\textwidth]{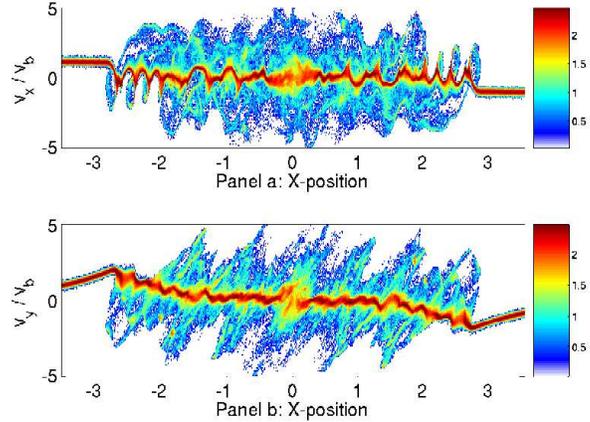}
\end{center}
\caption{The projected electron phase space distributions in run 1 at the time $t=56$: The panel (a) and (b) show the 10-logarithm of the $x,v_x$ and $x,v_y$ distributions, integrated over $0 < y < 0.1$ and in units of a computational electron. The electrons are accelerated in $v_x$ and $v_y$ by their interaction with the upper-hybrid waves and $\mathbf{B}_0$.}\label{fig2}
\end{figure}

Figure \ref{fig2} shows the corresponding electron phase space distribution.
The electrons have been accelerated in the interval $-2.8 < x < 2.8$, in which the $E_x$ component oscillates with a high amplitude. The electrons are modulated coherently in the $x,v_x$ plane only close to the respective cloud fronts. The modulation does not show the closed elliptic orbits, which are a characteristic of the saturated waves driven by the Buneman instability \cite{Schamel1} and which can initially be observed also for saturated upper-hybrid waves \cite{Paper15}. The phase speed of these waves equals the ion beam speed. The electrostatic potential, which traps the electrons, should thus move in the reference frame of the simulation box at $v_b$ for $x>0$ or $-v_b$ for $x<0$ \cite{Schamel1}. The orbits of the trapped electrons are, however, not centred at these speeds in Fig. \ref{fig2}(a).

Instead, the centres of the phase space structures in Fig. \ref{fig2}(a) move at $v \approx 0$. We attribute this and the absence of closed electron orbits to the magnetic field that accelerates the electrons in the $v_y$-direction in Fig. \ref{fig2}(b). This acceleration imposes a drag on the wave that is responsible for the reduced phase speed \cite{KrasSag}, explaining also that the wavelengths of the oscillations of $E_x$ in Fig. \ref{fig1} are shorter than $\lambda_u$. Note, however, that the electrons rapidly detrap and that they do thus not form the phase space cones \cite{Paper15} that evidence electron surfing acceleration. The phase space structures in the $x,v_x$ plane in the interval $-1.5 < x < 1.5$ show twice the length of those at the cloud fronts. They have been formed by the coalescence instability \cite{Berk}. Figure \ref{fig2}(b) also reveals that cusps in the $v_y$ speed of the electrons have developed at the leading edges of the clouds at $x \approx \pm 2.8$, which must be linked to a $B_z$ that varies along $x$. 

\begin{figure}[t]
\begin{center}
\includegraphics[width=0.5\textwidth]{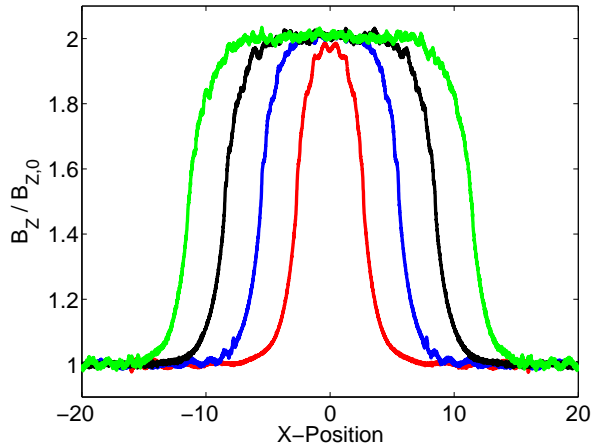}
\end{center}
\caption{The $B_z / B_{z,0}$ component sampled at $y=0$: The curves represent the time $t=56$ (red), $t=112$ (blue), $t=168$ (black) and $t=225$ (green). The magnetic field is increased from the initial value to twice that value in the cloud overlap region. The overlap region with the amplified $B_z$ expands in time, due to the ion beam propagation.}\label{fig3}
\end{figure}

The magnetic $B_z$ component is indeed not spatially uniform. Figure \ref{fig3} demonstrates that the value of the background magnetic field is amplified to $B_z / B_{z,0} = 2$ in the cloud overlap region. Snapshots at four simulation times evidence that the interval with $B_z \approx 2 \, B_{z,0}$ expands at a constant rate. This magnetic expansion is driven by the ion beams that do not change their speed much during the simulation time.
The magnetic field amplitude will increase beyond $2B_{z,0}$ when a shock starts to form, as PIC simulations of perpendicular shocks demonstrate \cite{Paper24,Paper25,Paper26}. 

\begin{figure}[t]
\begin{center}
\includegraphics[width=0.5\textwidth]{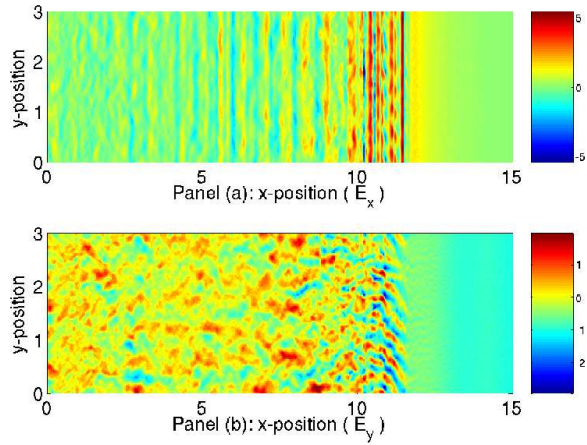}
\end{center}
\caption{The electric fields in simulation 1 at $t=225$: Panel (a) shows the $E_x$ component, which is due to the upper-hybrid wave with its $\mathbf{k} \parallel \mathbf{v}_b$. It shows planar structures. Panel (b) shows $E_y$. The oblique mixed modes and charge density modulations by electron bunches contribute to this electric field component.}\label{fig4}\end{figure}

The movie 1 shows that the electric field is predominantly polarized along the x-direction and that the field structures are planar before $t\approx 56$. After this time the $E_y$-field component starts to be structured along the y-direction, which is usually a characteristic of oblique modes and of filamentation modes. It can, however, also be caused by bunches of accelerated electrons that yield charge density fluctuations. Figure \ref{fig4} shows the $E_x$ and $E_y$ components at $t=225$. Both fields have amplitudes exceeding $E_c$. The $E_x$ is strongest and shows planar wave fronts, in particular close to the leading edge of the cloud at $x \approx 11$. Filamentary structures in $E_y$ are observed in the overlap region of both clouds, but their amplitude is lower. The strongest oscillations of $E_y$ reach $2 \, E_c$ at the leading edge. The $E_x$ oscillations are 2.5 times stronger than those of $E_y$. The movie 2 shows that the structuring of $E_y$ at the ion beam front along the y-direction changes only slowly in time. The oscillations of $E_y$ at $10<x<12$ in Fig. \ref{fig4}(b) are thus caused by an instability between the ion beam and the incoming plasma \cite{NewJPhys}. The structures for $x<8$ in Fig. \ref{fig4}(b) fluctuate according to the movie 2 and they are probably caused by fast electron bunches that move in the x-y plane. 

\begin{figure}[t]
\begin{center}
\includegraphics[width=0.5\textwidth]{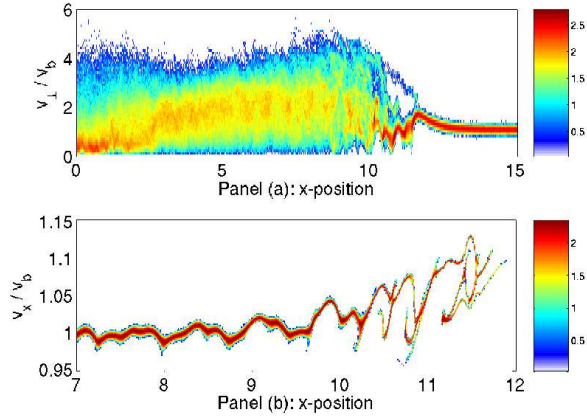}
\end{center}
\caption{The particle distributions in run 1 integrated over $0<y<0.1$ at $t=225$: Panel (a) displays the 10-logarithm of the electron distribution, as a function of $v_\perp$. The electrons are rapidly heated by the ion beam front. Panel (b) shows the 10-logarithmic ion beam distribution, which shows phase space holes at its front. Both panels show the phase space density in units of a computational particle.}\label{fig5}
\end{figure}

Figure \ref{fig5} shows the final phase space distributions of the electrons and ions. We compute the electron phase space density as a function of $x$ and $v_\perp = {(v_x^2 + v_y^2)}^{1/2}$ and the ion phase space density as a function of $x$ and $v_x$. The distributions are integrated over the interval $0 < y < 0.1$. We find that the electrons are accelerated to about $5 \, v_b$ (16 keV), which is 2.5 times the streaming speed of the ions relative to the incoming electrons. The momentum modulation of the ion beam is of the order of 1\% and rises to about 10\% at the cloud front.
The electrons in the overlap region show a spatially almost uniform velocity distribution. The ions, on the other hand, have not yet thermalized. 

\subsection{Collision of clouds with non-equal densities: run 2}

\begin{figure}[t]
\begin{center}
\includegraphics[width=0.5\textwidth]{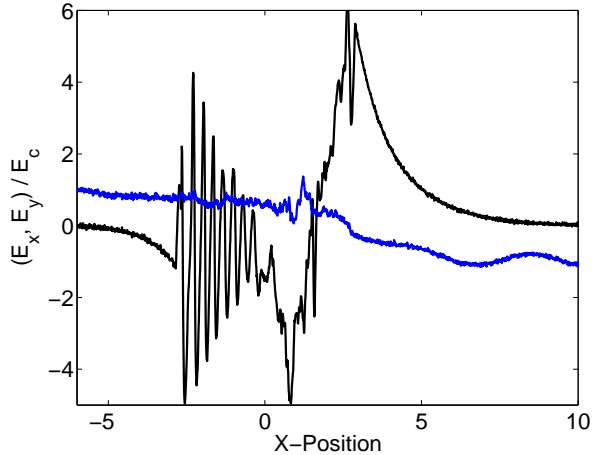}
\end{center}
\caption{The electric fields $E_x$ (black) and $E_y$ (blue) at the position 
$y=0$ and time $t=56$: The $E_x$ field is oscillatory in the interval 
$-3 < x < 0$. A single wave structure occurs in the $E_x$-field for 
$0<x<5$. Its oscillation period far exceeds $\lambda_u$.
The $E_y$ in this interval oscillates with a low amplitude around $E_y=0$.
\label{fig6}}
\end{figure}

Figure \ref{fig6} displays the electric $E_x$ and $E_y$ components at $t=56$.
The $E_x$-component behaves qualitatively different in the interval $x<0$, in which the tenuous ion beam interacts with the dense plasma of cloud 1 and in the interval $x>0$, where the dense ion beam interacts with the tenuous plasma of cloud 2. The oscillations for $x<0$ have a $\lambda \approx \lambda_u/2$, as in run 1. Only weak oscillations with $\lambda \approx \lambda_u / 2$ are observed at $x\approx 3$, which are superposed onto a powerful wave structure with $\lambda \gg \lambda_u$. This electric field forms a ramp for the ions. The ions of cloud 2, which move to decreasing values of $x$, will be slowed down in the interval $1 < x < 5$ and the ions of cloud 1 moving to increasing $x$ are decelerated in $0<x<1$. The movie 3 demonstrates that the large-scale structure is long-lived and that it does not develop a significant structuring in the y-direction. The movie 4 shows that this large-scale structure has no $E_y$ component.  

\begin{figure}[t]
\begin{center}
\includegraphics[width=0.5\textwidth]{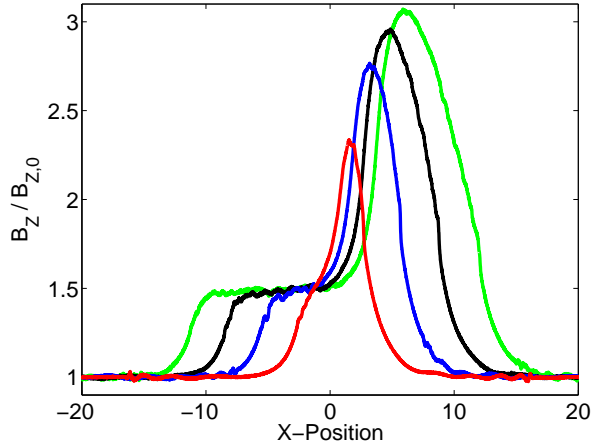}
\end{center}
\caption{The $B_z/B_{z,0}$ sampled at $y=0$: The curves represent the time $t=56$ (red), $t=112$ (blue), $t=168$ (black) and $t=225$ (green). The $B_z$ is amplified in the cloud overlap layer. The maximum of $B_z$ is located close to the front of the dense ion beam. It propagates with the beam and the maximum of $B_z$ increases with time.}\label{fig7}
\end{figure}

The large-scale electric field structure is correlated with a change in the strength of $B_z$. The $B_z$ has been compressed during the time interval up to $t=56$. Figure \ref{fig7} demonstrates this for $y=0$ sampled at four times.
The $B_z$ at $t=56$ peaks at $x\approx 1.2$, which coincides with the position where the $E_x$ of the large-scale electric field structure in Fig. \ref{fig6} changes its sign. A comparison of $E_x$ with $B_z$ at $t=56$ suggests that $E_x$ changes with $dB_z / dx$. This can be interpreted as $E_x \propto B_z (dB_z / dx)$, since $B_z > 0$ everywhere. The exact proportionality cannot be demonstrated, because short waves are superposed on the large-scale structure. A proportionality of $E_x \propto B_z (dB_z / dx)$ would imply that the large-scale $E_x$ is driven by the magnetic pressure gradient. Figure \ref{fig7} also reveals that the amplification of $B_z$ is lower for $x<0$ in comparison to run 1.

\begin{figure}[t]
\begin{center}
\includegraphics[width=0.5\textwidth]{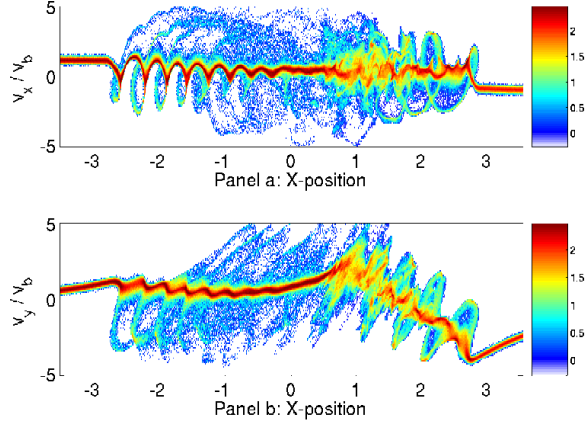}
\end{center}
\caption{The electron phase space distributions in run 2 at $t=56$: Panel (a) and (b) show the 10-logarithm of the $x,v_x$ and $x,v_y$ distributions, integrated over $0 < y < 0.1$ and in units of a computational electron of the dense cloud. The electrons are accelerated by their interaction with the upper-hybrid wave and with $\mathbf{B}_0$.}\label{fig8}
\end{figure}

Figure \ref{fig8} displays the electron distribution in run 2 at $t=56$. The $x,v_x$ distribution shows well-defined electron phase space holes that involve the bulk of the electrons, but only at the leading edge of the tenuous cloud at $x \approx -2$. 
The periodic structures in $x,v_x$ are apparently stable only in the interval, in which the $v_y$ speed of the bulk electrons and thus $B_z$ does not significantly change. The rapid change in the mean $v_y$ component at $x \approx 1.5$ destroys the electron phase space holes and the bulk electrons are heated. Some oscillatory electron orbits are visible for $x>0$ but the number density of the electrons following these paths is low. The lack of periodic electron structures with a significant charge density explains, why no strong oscillations in $E_x$ on a scale $\approx \lambda_u$ are visible in this interval in Fig. \ref{fig6}. Like in Fig. \ref{fig2}(a) the electron phase space holes in Fig. \ref{fig8}(a) are tied to an electrostatic wave with a phase speed, which is lower than that driven by the Buneman- or the upper-hybrid wave instability.

\begin{figure}[t]
\begin{center}
\includegraphics[width=0.5\textwidth]{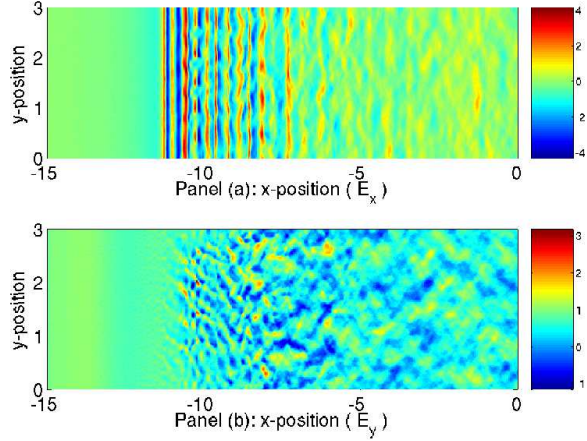}
\end{center}
\caption{The electric fields in simulation 2 for $x<0$: Panel (a) shows the $E_x$ component and panel (b) the $E_y$ component at $t=225$. The planar structures of $E_x$ at $x<-10$ are due to the upper-hybrid waves. These waves damp out for increasing $x>-10$ and they loose their planarity. The typical size of the structures of $E_y$ increase for increasing $x>-10$ and their amplitude decreases.}\label{fig9}
\end{figure}

Figure \ref{fig9} shows the $E_x$ and $E_y$ fields in the half-plane $x<0$ at $t=225$. The amplitude of $E_x$ is about twice that of $E_y$ and it shows small scale, quasi-planar oscillations. These oscillations are characteristic for electron phase space holes that are extended along the y-direction in form of tubes. The $E_y$ field shows fluctuations along both directions. As we move away from the beam front at $x \approx -11$ to increasing values of $x$, the characteristic size of the fluctuations in $E_y$ increase and the structures in $E_x$ become less planar. The $E_x$- and $E_y$ amplitudes are lower than their counterparts in Fig. \ref{fig4}, but their spatial distribution is similar.

\begin{figure}[t]
\begin{center}
\includegraphics[width=0.5\textwidth]{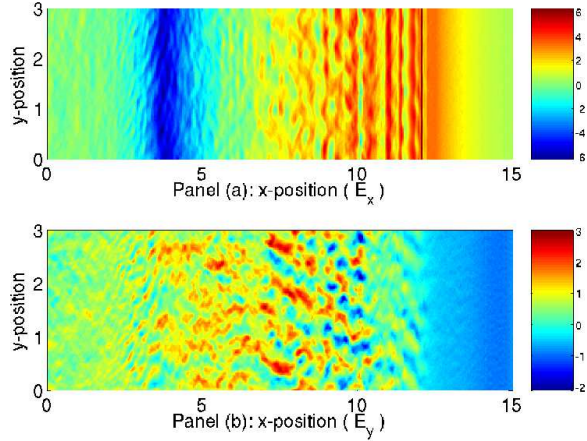}
\end{center}
\caption{The electric fields in simulation 2 for $x>0$ at $t=225$: Panel (a) shows the $E_x$ component and panel (b) the $E_y$ component. Planar wave structures in $E_x$ are superposed on a large-scale electric field modulation. The $E_y$ component shows structuring along $y$ at $x\approx 9$, which decays for decreasing values of $x$.}\label{fig10}
\end{figure}

Figure \ref{fig10} shows the $E_x$ and $E_y$ in the half-plane $x>0$. The small-scale oscillations of $E_x$ are comparable to those of $E_y$. A planar negative electric field structure is found at $x=4.5$ and the oscillations close to the cloud front at $x=11$ are superposed on a structure with a positive $E_x$. The $E_y$ fields are strong only in the interval $5<x<10$. The large-scale fields exceed the oscillations in $E_x$ and $E_y$ in their amplitude and should be the strongest accelerators in the plasma system, due to their spatial coherence over more than one $\lambda_s$. The field structure is also stable, as the movie 3 evidences.
This structure is again related to the steepest gradients of $B_z$ at $t=225$ in the Fig. \ref{fig7}.

\begin{figure}[t]
\begin{center}
\includegraphics[width=0.5\textwidth]{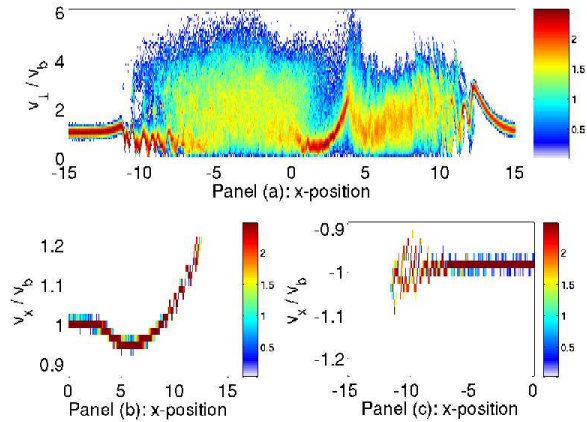}
\end{center}
\caption{The particle distributions in run 2 integrated over $0<y<0.1$ at $t=225$: Panel (a) displays the 10-logarithm of the electron distribution, as a function of $v_\perp = {(v_x^2+v_y^2)}^{1/2}$. Panels (b) and (c) show the 10-logarithmic phase space distributions of the respective beams. Both panels show the phase space density in units of a computational particle of the dense plasma cloud. The electrons have been thermalized for $-8<x<0$ and $5<x<10$ and a beam exists at $x\approx 4$. Only the ions of the dense beam in (b) are significantly accelerated.}\label{fig11}
\end{figure}

Figure \ref{fig11} shows the final electron and ion distributions in run 2, which are both asymmetric relative to $x=0$. The electron acceleration in $x<0$ is somewhat stronger. The electron peak speed is comparable to that in run 1 in Fig. \ref{fig5}. The accelerating electron beams in the intervals $0<x<5$ and $12<x<15$ are related to the magnetic field gradients in Fig. \ref{fig7} and involve the $v_y$ component. The ion beam going to positive $x$ in Fig. \ref{fig11}(b) confirms what we anticipated in the discussion of Fig. \ref{fig10}. The ions slow down in the negative electric field at $x \approx 5$ and accelerate at $x \approx11$. The ions are accelerated by more than 20\%, which exceeds the ion acceleration in run 1. Figure \ref{fig11}(c) shows that the tenuous ions going to negative $x$ are almost un-accelerated. The ions spread out at the cloud front due to their electrostatic interaction with the electrons.

\section{Discussion}

In this work we have examined with particle-in-cell (PIC) simulations the collision of two magnetized plasma clouds. Plasma instabilities form in the layer, in which both clouds overlap. After a short transient phase the plasma thermalization is accomplished by an ion beam moving through a cool electron plasma. The cool electron plasma is from that cloud that has not yet interacted with the second cloud. The electrons of the second cloud have been heated. They are confined by the perpendicular magnetic field to the layer, in which both clouds overlap. This layer expands. One ion beam-driven instability develops at each of the two fronts of the overlap layer. 

The plasma clouds have been equally dense in one simulation, which is typically assumed by PIC simulations of kinetic shocks that employ the piston method \cite{Paper24,Paper25,Paper26}. The piston method models one plasma cloud, which it reflects at one simulation boundary onto itself. Two clouds are modelled here and we get two instabilities with equally dense ion beams, one at each front of the expanding overlap layer. The ion beam density is comparable to that of the electron beam, which is the equivalent of the Buneman instability in magnetized plasma. The density ratio has been set to 2 in the second simulation, which gives one instability driven by a dense ion beam and one driven by a tenuous ion beam. We could thus assess the impact of the ion beam density on the plasma dynamics for three cases using two simulations. 

The simulation parameters are probably relevant for the foreshock regions of the fastest perpendicular shocks at supernova remnants (SNRs). The normalization of the Maxwell-Lorentz set of equations makes the results independent of the value of the plasma frequency in physical units. The strength of the magnetic field resulted in a ratio of 5 between the electron plasma frequency of the dense cloud and the electron cyclotron frequency. The equivalent value close to SNR shocks is thought to be several hundred. However, a substantial decrease of this ratio to about 10 might occur at some SNR shocks \cite{VolkM}. Shock-accelerated particles can also amplify the magnetic field locally \cite{BellM}. We have selected a ratio of 5, because it allows us to resolve magnetic field effects such as the electron surfing acceleration \cite{Paper15}.

Some upstream ions are reflected by the shock [6-10] and they gyrate in the foreshock, which is a scenario analogous to that we have modelled here. The ion beam speed can reach twice the shock speed in the upstream frame of reference, if the reflection is specular. Our ion beam speed in the reference frame of the electron beam is 0.1 c, which can be achieved with realistic SNR shock speeds \cite{Kulkarni,FryerNew}. It is between the low speeds, for which electrostatic instabilities dominate, and the fast speeds that favor mixed- or filamentation modes \cite{Paper13,NewJPhys}. Our simulations here have shown that the wave structures are quasi-planar and the electrostatic instabilities thus dominate. The emergence of filamentary structures indicates, however, that this speed is probably the most extreme case in this respect, at least for our plasma parameters.

We have introduced a second spatial dimension and an ion beam with a finite length. This has left unchanged the findings of previous one-dimensional PIC simulations of spatially uniform beam distributions with respect to the acceleration of electrons. The peak speed of the electrons is still limited to a few times the ion beam speed in the reference frame of the electrons. This peak energy has furthermore been independent of the ion beam density. The peak energy has been 10-20 keV for the ion beam speed 0.1 c, which falls short of the 100 keV that are required for the electron injection into diffusive acceleration \cite{Paper4}. Higher electron energies may be reached when the ions thermalize.

If both plasma clouds have the same density, the ion acceleration is identical close to the fronts of both beams and it amounts to maximally $v_b / 10$. The simulations have shown that the magnetic field is convected with the plasma slabs, as expected from the frozen-in theorem. The magnetic field strength doubles where both clouds overlap. We could observe well-developed electron phase space holes. The speed of their electrostatic potentials has been low, presumably due to the magnetic drag \cite{KrasSag} excerted on the trapped electrons. The electron surfing acceleration \cite{Paper15} by the transport of the trapped electrons across the magnetic field has been weak. 

The ions at the front of the dense beam have been accelerated by $v_b/5$ in the simulation with the unequal cloud densities, while those at the front of the tenuous beam have only been scattered. This has originated in the asymmetric magnetic field compression. The magnetic field strength has increased by only 50\% at the front of the tenuous beam while it has reached over 3 times its initial value at the front of the dense beam. The magnetic field strength has also varied more rapidly. This variation was correlated with a large-scale electric field, which may have been driven by the magnetic pressure gradient. It could strongly interact with the ions. This electric field has destroyed the electron phase space holes in the bulk plasma, probably because it has removed the stable equilibrium position of the potential.

{\bf Acknowledgments:} The European Social Fund and National Resources (EPEAEK II) PYTHAGORAS, the German DFG, the Irish DIAS and the Swedish Vetenskapsr\aa det and computer centre HPC2N (Ume\aa) have supported this project.

\end{document}